\documentclass{elsart}

\usepackage{graphicx}
\usepackage{amsmath}
\usepackage{natbib}
\usepackage{booktabs}

\begin{document}

\begin{frontmatter}

\title{A Time-Symmetric Block Time-Step Algorithm for N-Body Simulations}

\author{Junichiro Makino\thanksref{1}},
\author{Piet Hut\thanksref{2}},
\author{Murat Kaplan\thanksref{3}}, 
\author{Hasan Sayg\i n\thanksref{4}}

\thanks[1]{Department of Astronomy, University of Tokyo, 7-3-1 Hongo,
  Bunkyo-ku, Tokyo 113-0033, Japan}
\thanks[2]{Institute for Advanced Study, Princeton, NJ 08540, USA}
\thanks[3]{Istanbul Technical University, Informatics Institute, Computational 
Science and Engineering Program,  34469 Maslak, Istanbul, Turkey}
\thanks[4]{Istanbul Technical University, Institute of Energy, 34469 Maslak, 
Istanbul, Turkey}

\begin{abstract}
The method of choice for integrating the equations of motion of the
general N-body problem has been to use an individual time step scheme.
For the sake of efficiency, block time steps have been the most popular,
where all time step sizes are smaller than a maximum time step size by
an integer power of two.  We present the first successful attempt to
construct a time-symmetric integration scheme, based on block time steps.
We demonstrate how our scheme shows a vastly better long-time behavior
of energy errors, in the form of a random walk rather than a linear drift.
Increasing the number of particles makes the improvement even more pronounced.
\end{abstract}

\end{frontmatter}
\maketitle

\section{Introduction}

For long-term N-body simulations, it is essential that the drift in
the values of conserved quantities is kept to a minimum.  The total
energy is often used as an indicator of such a drift.  During the
last fifteen years, two approaches have been put forward to improve
numerical conservation of energy and other theoretically conserved
quantities: symplectic integration schemes, where the simulated system
is guaranteed to follow a slightly perturbed Hamiltonian system, and
time-symmetric integration schemes, where the simulated system follows
the same trajectory in phase space, when run backward or forward

In both cases, for symplectic as well as for time-symmetric schemes,
the introduction of adaptive time steps tends to destroy the desired
properties.  Symplectic schemes are perturbed to different Hamiltonians
at different choices of time step length, and therefore lose their
global symplecticity.  Time-symmetric schemes typically determine
their time step length at the beginning of a step, which implies that
running a step backward gives a slightly different length for that step.
See \citet{Leimkuhler-2005} for a recent review of various attempts to
remedy this situation.

In practice, many large-scale simulations in stellar dynamics use a
block time-step approach, where the only allowed values for the time
step length are powers of two \citep{Aarseth-2003}.  The name derives
from the fact that, with this recipe, many particles will share the
same step size, which implies that their orbit integration can be
performed in parallel.  An added benefit, in the case of individual
time steps, is the fact that block time steps allow one to predict
the positions of all particles only once per block time step, rather
than separately for each particle that needs to be moved forward.
Since parallelization is rapidly becoming essential for any major
simulation, we explore in this paper the possibility to extend time
symmetry to the use of block time steps.

In Section 2, we analyze some of the problems that occur when applying
existing methods to the case of block time steps, and we offer a novel
solution, with a truly time symmetric choice of time step, with the
restriction that we only allow changes of a factor two in the
direction of increasing and decreasing the time step.
In Section 3, we present numerical tests of our new scheme in the
simplest case of the 2-body problem, which already shows the
superiority of our approach over various alternatives.  In Section 4,
we demonstrate that the advantage carries over to the general N-body
problem.  Section 5 sums up.

\section{Block Time Steps}

\subsection{Implicit Iterative Time Symmetrization}
\label{subsec:implicit}

It is surprisingly easy to introduce a time-symmetric version for any
adaptive self-starting integration scheme.  Let $\xi = (r,v)$ be the
$2N$-dimensional phase space vector for a system with $N$ degrees of
freedom, and let $f(\xi_i,\delta t_i)$ be the operator that maps the
phase space vector of the system at time $t_i$ to a new phase space
vector at time $t_{i+1} = t_i + \delta t_i$.  Any choice of
self-starting integration scheme, together with a recipe to determine
the next time step $\delta t_i$, at time $t_i$ and phase space value
$\xi_i(t_i)$, defines the precise form of $f(\xi_i,\delta t_i)$.

The recipe for making any such scheme time-symmetric was given by
\citet{Hut-1995}, as:


\begin{equation}
\left\{ \begin{array}{lcl}
\xi_{i+1} &=& f(\xi_i,\delta t_i),\\
\phantom{1}&\phantom{1}&\phantom{1} \\
\delta t_i &=&
{\displaystyle \frac{h(\xi_i)+h(\xi_{i+1})}{2}} \label{time_symm}
\end{array} \right.
\end{equation}

where $h(\xi)$ can be any time step criterion.  Note that this recipe
leads to an implicit integration scheme, which can be solved most
easily through iteration.  In practice, one or two iterations suffice
to get excellent accuracy, but at the cost of doubling or tripling the
number of force calculations that need to be performed.  Extensions
of this implicit symmetrization idea have been presented by
\citet{Funato-1996} and \citet{Hut-1997}.

Since we will need to inspect the idea of iteration below in more detail,
let us write out the process here explicitly.  We start with the given
state $\xi_i$ and the implicit equation for $\xi_{i+1}$ of the form

\begin{equation}
\xi_{i+1} = f(\xi_i,\delta t_i(\xi_i, \xi_{i+1}))
\end{equation}

The first guess for $\xi_{i+1}$ is

\begin{equation}
\xi_{i+1}^{(0)} = f(\xi_i,\delta t_i(\xi_i, \xi_i))
\end{equation}

and we can consider this as our zeroth-order iteration.  With this
guess in hand, we can now start to iterate, finding

\begin{equation}
\xi_{i+1}^{(1)} = f(\xi_i,\delta t_i(\xi_i, \xi_{i+1}^{(0)}))
\end{equation}

as our first-order iteration.  This will already be much closer to the
final value, as long as the time steps are small enough and the
function $\delta t_i$ does not fluctuate too rapidly.  In general, the
$k^{th}$ iteration will yield a value for $\xi_{i+1}$ of

\begin{equation}
\xi_{i+1}^{(k)} = f(\xi_i,\delta t_i(\xi_i, \xi_{i+1}^{(k-1)}))
\end{equation}

We will now consider the application of these techniques to block time
steps.  For the purpose of illustrating the use of block time steps,
it will suffice to use the leapfrog scheme (also known as the
Verlet-St\"ormer-Delambre scheme, according to the authors who
rediscovered this scheme at roughly century-long intervals), which we
present here in a self-starting, but still time-symmetric form:

\begin{eqnarray}
r_{i+1}&=& r_{i} + v_{i}\delta t + a_i(\delta t)^2/2 \nonumber\\
v_{i+1} &=& v_i + (a_i + a_{i+1}) \delta t/2	 \label{leapfrog}
\end{eqnarray}

All our considerations carry over to higher-order schemes, as long as
the base scheme can be made time-symmetric when iterated to
convergence.  An example of such a scheme is the widely used Hermite
scheme\citep{Makino-1991}.

\subsection{Flip-Flop Problem}
\label{subsec:flipflop}

To start with, we apply the recipe of \citet{Hut-1995} to block time steps.
Let us define a block time step at level $n$ as having a length:

\begin{equation}
\Delta t_n = \frac{\Delta t_1}{2^{n-1}}. \label{block_time}
\end{equation}

where $\Delta t_1$ is the maximum time step length.
Starting with the continuum choice of

\begin{equation}
\delta t_{c,i} = \frac{h(\xi_i)+h(\xi_{i+1})}{2}
\end{equation}

we now force each time step to take on the block value
$\delta t_i = \Delta t_n$ for the smallest $n$ value
that obeys the condition
$\Delta t_n \le \delta t_{c,i}$.
In more formal terms,
$\delta t_i = \delta t_i(\delta t_{c,i}) = \Delta t_n$
for the unique $n$ value for which

\begin{equation}
n = \min_{k \ge 1}
\left\{
 k \ \Big| \ \frac{1}{2^{k-1}} \le \frac{h(\xi_i)+h(\xi_{i+1})}{2}
\right\}                                           \label{blockcondition}
\end{equation}

The problem with this approach is that we are no longer guaranteed to
find convergence for our iteration process, as can be seen from the
following example.  Let $h(\xi_i) = 0.502$ and let the time derivative
of $h(\xi_i(t))$ along the orbit be $(d/dt)h(\xi_i(t)) = -0.01$.  We
then get the following results for our attempt at iteration.

\begin{eqnarray}
\xi_{i+1}^{(0)} &=& f(\xi_i,\delta t_i(\xi_i, \xi_i))
= f(\xi_i,\delta t_i(h(\xi_i)))                        \nonumber\\
&=& f(\xi_i,\delta t_i(0.502))     
= f(\xi_i, 0.5)
\end{eqnarray}

\begin{eqnarray}
\xi_{i+1}^{(1)} &=& f(\xi_i,\delta t_i(\xi_i, \xi_{i+1}^{0}))
= f(\xi_i,\delta t_i([h(\xi_i) + h(\xi_{i+1}^{0})]/2))   \nonumber\\
&=& f(\xi_i,\delta t_i([0.502 + (0.502+0.5*(-0.01))]/2))         \nonumber\\
&=& f(\xi_i,\delta t_i([0.502 + 0.497]/2))                       \nonumber\\
&=& f(\xi_i,\delta t_i(0.4995))       
= f(\xi_i, 0.25)
\end{eqnarray}

\begin{eqnarray}
\xi_{i+1}^{(2)} &=& f(\xi_i,\delta t_i(\xi_i, \xi_{i+1}^{1}))
= f(\xi_i,\delta t_i([h(\xi_i) + h(\xi_{i+1}^{1})]/2))        \nonumber\\
&=& f(\xi_i,\delta t_i([0.502 + (0.502+0.25*(-0.01))]/2))         \nonumber\\
&=& f(\xi_i,\delta t_i([0.502 + 0.4995]/2))                       \nonumber\\
&=& f(\xi_i,\delta t_i(0.50075))        
= f(\xi_i, 0.5)
\end{eqnarray}

And from here on, $\xi_{i+1}^{(k)} = f(\xi_i, 0.25)$ for every odd value of $k$
and $\xi_{i+1}^{(k)} = f(\xi_i, 0.5)$ for every even value of $k$: the process
of iteration will never converge.

Under realistic conditions, for slowly varying $h$ functions and small
time steps, this flip-flop behavior will not occur often, but it will
occur sometimes, for a non-negligible fraction of the time.  We can
see this already from the above example: for a linear decrease in the
$h$ function of $(d/dt)h(\xi_i(t)) = -0.01$, we will get flip-flopping
not only for $h(\xi_i) = 0.502$ but for any value in the finite range
$ 0.50125 < h(\xi_i) < 0.5025$.

Since iteration converges correctly over the rest of the interval
$ 0.5 < h(\xi_i) < 1$, we conclude that in this particular case
flip-flopping occurs about one quarter of one percent of the time,
over this interval.  This is far too frequent to be negligible in a
realistic situation.

Clearly, a straightforward extension of the implicit iterative time
symmetrization approach does not work for block time steps, because
iteration does not converge.  We have to add some feature, in some
way.  Our first attempt at a solution is to take the smallest of the
two values in a flip-flop situation.

\subsection{Flip-Flop Resolution}
\label{subsec:noflipflop}

The most straightforward solution of the flip-flop dilemma is like
cutting the Gordian knot: we just take the lowest value of the two
alternate states.  The drawback of this solution is that in general
we need at least two iterations for each time step, to make sure that we
have spotted, and then correctly treated, all flip-flop situation.
In general, it is only at the third iteration that it becomes obvious
that a flip-flop is occurring.  To see this, consider the previous
example with a starting value of $h(\xi_i) = 0.501$.
In that case we will get $\xi_{i+1}^{(0)} = f(\xi_i, 0.5)$ and
$\xi_{i+1}^{(1)} = f(\xi_i, 0.25)$, just as when we started with
$h(\xi_i) = 0.502$.
The difference shows up only at the second iteration, where we now
find $\xi_{i+1}^{(2)} = f(\xi_i, 0.25)$, a value that will hold for
all higher iterations as well.

The original iterative approach to time symmetry in practice already
gives good results when we use only one iteration.  This implies a
penalty, in terms of force calculations per time steps, of a factor
two compared to non-time-symmetric explicit integration.  Now the
use of flip-flop resolution will force us to always take at least two
iterations per step, raising the penalty to become at least a factor
of three.

However, there is a more serious problem: there is still no guarantee
that taking the lowest value in a flip-flop situation leads to a
time-symmetric recipe.  In fact, what is even more important, we have
not yet checked whether our symmetric block time-step scheme is really
time symmetric, in the absence of flip-flop complications.

In order to investigate these questions, let us return to the example
we used above, but instead of a linear time derivative, let us now use
a quadratic time derivative for the $h$ function that gives the
estimate for the time step size.  Rather than writing a formal
definition, let us just state the values, while shifting the time
scale so that $t=0$ coincides with the particle position being
$\xi_i$:

\begin{eqnarray}
h(0.00) &=& 0.502 \nonumber\\
h(0.25) &=& 0.499 \nonumber\\
h(0.50) &=& 0.499
\end{eqnarray}

When we start at time $t=0$, and we integrate forward, we find:

\begin{eqnarray}
\xi_{i+1}^{(0)} &=& f(\xi_i,\delta t_i(\xi_i, \xi_i))
= f(\xi_i,\delta t_i(h(\xi_i)))                        \nonumber\\
&=& f(\xi_i,\delta t_i(0.502))     
= f(\xi_i, 0.5)
\end{eqnarray}

\begin{eqnarray}
\xi_{i+1}^{(1)} &=& f(\xi_i,\delta t_i(\xi_i, \xi_{i+1}^{0}))
= f(\xi_i,\delta t_i([h(\xi_i) + h(\xi_{i+1}^{0})]/2))   \nonumber\\
&=& f(\xi_i,\delta t_i([0.502 + 0.499]/2))                       \nonumber\\
&=& f(\xi_i,\delta t_i(0.5005))       
= f(\xi_i, 0.5)
\end{eqnarray}

and so on: all further $k$th iterations will result in
$\xi_{i+k}^{(1)} = f(\xi_i, 0.5)$.  There is no flip-flop situation,
when moving forward in time.

However, when we now turn the clock backward, after taking this step
of half a time unit, we start with the value $h(0.50) = 0.499$, which
leads to a first step back of $\delta t = 0.25$.  The end point of the
first step back is $t = 0.25$ with $h(0.25) = 0.499$.  Therefore, also
here there is no flip-flop situation: all iterations, while going
backward, result in a time step size of $\delta t = 0.25$.

We have thus constructed a counter example, where forward integration
would proceed with time step $\delta t = 0.5$ and subsequent backward
integration would proceed with time step $\delta t = 0.25$.  Clearly,
our scheme is not yet time symmetric, even in the absence of a
flip-flop case.

\subsection{A First Attempt at a Solution}
\label{subsec:firsttry}

Let us rethink the whole procedure.  The basic problem has been that
the very first step in any of our algorithms proposed so far has not
been time symmetric.  The very first step moves forward, and leads to
a newly evolved system at the end of the first step.  Only {\it after}
making such a trial integration, do we look back, and try to restore
symmetry.  However, as we have seen, the danger is large that this
trial integration is not exhaustive: it may already go too far, or not
far enough, and thereby it may simply overlook a type of move that the
same algorithm would make if we would start out in the time-reversed
direction.

Formulating the problem in this way, immediately suggests a solution.
At any point in time, let us first try to make the largest step that
is allowed.  If that step turns out to be too large for our algorithm,
we try a step that is half that size.  if that step is too large still,
we again half the size, and so on, until we find a step size that
agrees with our algorithm, {\it when evaluated in both time directions.}
A similar treatment has been described by Quin et al (1997).

This type of approach is clearly more symmetric than what we have
attempted so far.  Instead of using information of the physical system
at the starting point of the next integration step, we only use a
mathematical criterion to find the largest time step size allowed at
that point, {\it and we then apply the physical criteria symmetrically
in both directions.}

Let us give an example.  If the largest time step size is chosen to be
unity, then at time $t = 0$ we start by considering this time step.
We try, in this order $\delta t = 1$, $\delta t = 0.5$, $\delta t = 0.25$,
and so on, until we find a time step for which integration starting in
the forward direction, and integration starting in the backward direction,
both result in the new time step being acceptable.  Let us say that
this is the case for $\delta t = 0.125$.

After taking this step, we are at time $t = 0.125$.  The largest time
step allowed at that point, forward or backward, is $\delta t = 0.125$.
Any larger time step would result in non-alignment of the block time
steps: in the backward direction it would jump over $t = 0$.  So at
this point we start by considering once more $\delta t = 0.125$.  If
that time step is too large, we try half that time step, halving it
successively until we find a satisfactory time step size.

Imagine that the second time step size is also $\delta t = 0.125$.  In
that case, we land at $t = 0.25$.  From there on, the maximum allowed
time step size is $\delta t = 0.25$, so the first try should be that
size.

In principle, this approach seems to be really time symmetric.
However, there is a huge problem with this type of scheme, as we have
just formulated it.  Imagine the system to crawl along with time steps
of, say $\delta t = 1/1024$, and reaching time $t = 1$.  Our new recipe
then suggests to start by trying $\delta t = 1$, a 1024-fold increase
in time step!  Whatever subtle physical effect it was that forced us
to take such small time steps, is completely ignored by the mathematical
recipe that forces us to look at such a ridiculously large time step.

For example, in the case of stellar dynamics, a double star may force
the stars that orbit each other to take time steps that are necessarily
far shorter than the orbital period.  Starting out with a trial step size
that is far larger than an orbital period may or may not give spuriously
safe-looking results.  Clearly, we have to exclude such enormous jumps in
time step.

\subsection{A Second Attempt at a Solution}
\label{subsec:secondtry}

The simplest solution to taming sudden unphysical increases in time
steps is to allow at most an increase of a factor two, in either the
forward or the backward direction.  This then implies that we can only
allow decreases of a factor two, and not more than two, in either
direction.  The reason is that a decrease of a factor four in one
direction in time would automatically translate into an increase of a
factor four in the other direction.

Note that we have to be careful with our time step criterion.  If we
allow time steps that are too large, we may encounter situations where
our time step criterion would suggest us to shrink time steps by a
factor of four, from one step to the other.  Since our algorithm does
not allow this, we can at most shrink by a factor of two, which may
imply an unacceptably large step.  However, if our time step criterion
is sufficiently strict, allowing only reasonably small time steps too
start with, it will be able to resolve the gradients in the criterion
in such as way as to handle all changes gracefully through halving and
doubling.

When we apply this restriction to the scheme outlined in the previous
subsection, we arrive at the following compact algorithm.

First a matter of notation.  Any block time step, of size $\delta t = 1/2^k$,
connects two points in time, only one of which can be written at
$t = Z/2^{(k-1)}$, with $Z$ an integer.  Let us call that time value an
{\it even time}, from the point of view of the given time step size, and
let us call that other time value an {\it odd time}.  To give an example, if
$\delta t = 0.125$, than $t = 0, 0.25, 0.5, 0.75, 1$ are all even times,
while $t = 0.125, 0.375, 0.625, 0.875$ are all odd times.

Here is our algorithm:

\begin{itemize}
\item[]
When we start in a given direction direction in time, at a given point
in time, we should determine the time step size of the last step made
by the system.  In that way, we can determine whether the current time
is even or odd, with respect to that last time step.

\medskip

\item[]
If the current time is odd, our one and only choice is: to continue
with the same size time step, or to halve the time step.  First, we
try to continue with the same time step.  If, upon iteration, that
time step qualifies according to the time-symmetry criterion used
before, Eq. \ref{blockcondition}, we continue to use the same time
step size as was used in the previous time step.  If not, we use half
of the previous time step.

\medskip

\item[]
If the current time step is even, we have a choice between three
options for the new time step size: doubling the previous time step
size, keeping it the same, or halving it.  We first try the largest
value, given by doubling.  If Eq.\ref{blockcondition} shows us that
this larger time step is not too large, we accept it, otherwise we
consider keeping the time step size the same.  If Eq.\ref{blockcondition}
shows us that keeping the time step size the same is okay, we accept
that choice, otherwise we just halve the time step, in which case no
further testing is needed.
\end{itemize}

Note that in this scheme, we always start with the largest possible
candidate value for the time step size.  Subsequently, we may consider
smaller values, but the direction of consideration is always from
larger to smaller, never from smaller to larger.  This guarantees that
we do not run into the flip-flop problem mentioned above.

\section{Numerical Tests for the 2-Body Problem}

\begin{figure}[ht]
\centering
\includegraphics[width=0.90\textwidth]{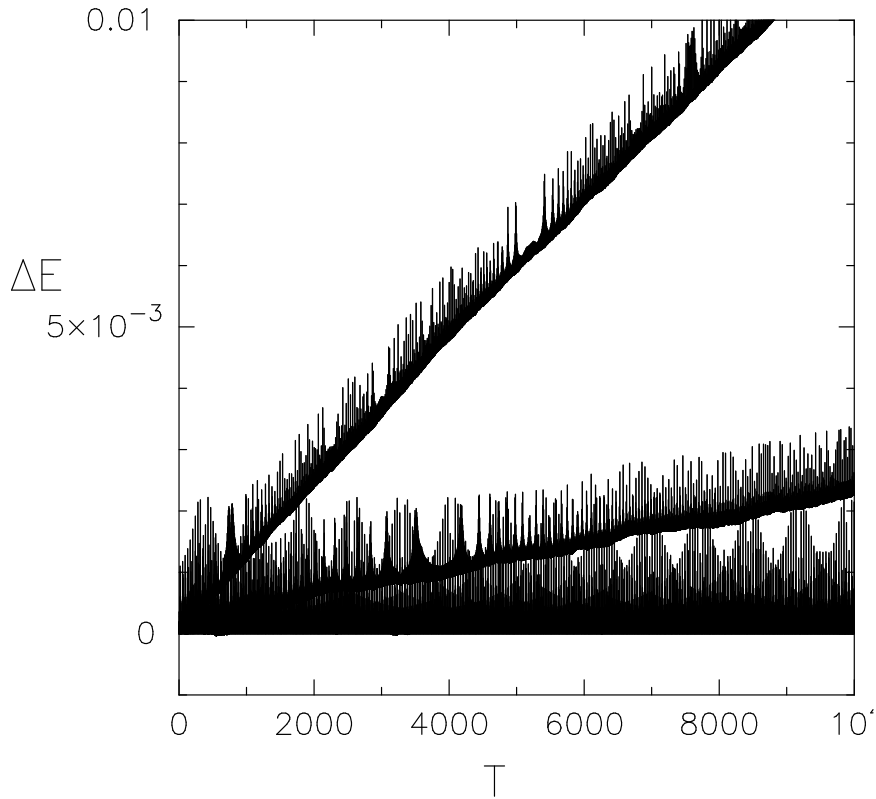}
\caption{Relative energy errors for a two-body integration of a bound orbit
  with eccentricity $e=0.99$.  The top line with highest slope
  corresponds to algorithm 1, the line with intermediate slope
  corresponds to algorithm 2, and below those the two lines for
  algorithms 0 and 3 are indistinguishable in this figure.}
\label{fig1}
\end{figure}

\begin{figure}[ht]
\centering
\includegraphics[width=0.90\textwidth]{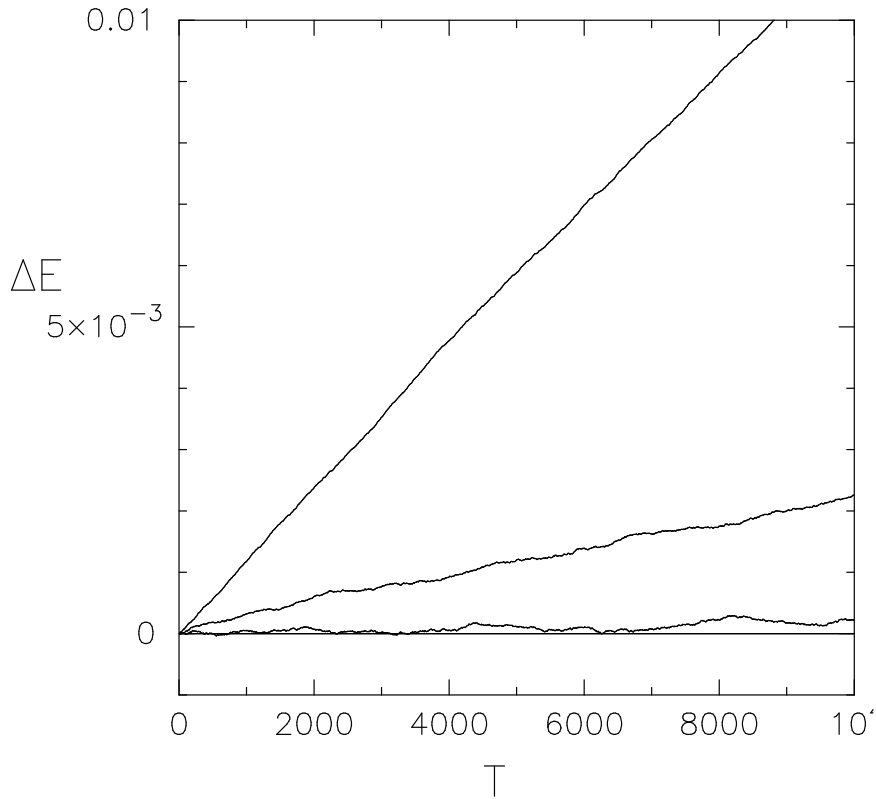}
\caption{Relative energy errors at apocenter.  The four lines, from top
to bottom, correspond to algorithms 1, 2, 3, and 0.}
\label{fig2}
\end{figure}

\begin{figure}[ht]
\centering
\includegraphics[width=0.90\textwidth]{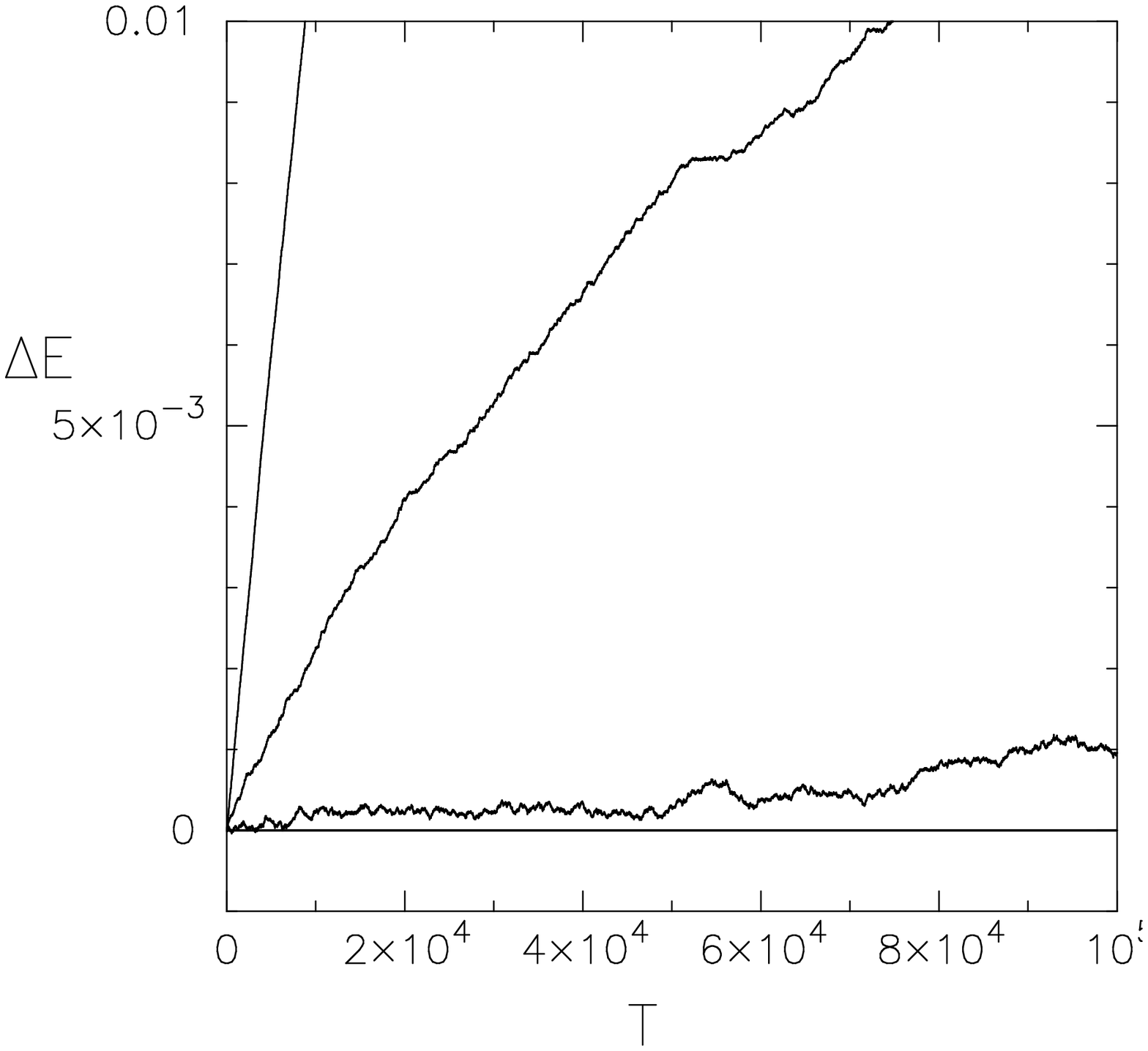}
\caption{Same as Fig. \ref{fig2}, but for a duration that is ten
times longer.}
\label{fig3}
\end{figure}

We present here the results for a gravitational two-body integration.
The relative orbit of the two point masses forms an ellipse with an
eccentricity of $e = 0.99$.  We have chosen a time unit such that the
period of the orbit is $T = 2\pi$.

We have implemented four different integration schemes:

\noindent
0) the original time-symmetric integration scheme described by
\citet{Hut-1995}, where there is a continuous choice of time step size.
This is the approach described in section \ref{subsec:implicit}.
We have used five iterations for each step.

\noindent
1) a block-time-step generalization, with a fixed number of iterations.
This is the approach analyzed in section \ref{subsec:flipflop}.  Here,
too, we chose five iterations for each step.

2) a block time step generalization, with a variable number of iterations.
If after five iterations, the fourth and the fifth iterations still give
a different block time step size, then we choose the smallest of the two.
This recipe avoids flip-flop situations.  It is the approach described
in section \ref{subsec:noflipflop}.

The algorithm described in the next section, \ref{subsec:firsttry}, we
have not implemented here, because it is guaranteed to lead to large
errors in those cases where a new large time step is allowed again
just before pericenter passage.  We therefore switched directly to the
following section:

3) the implementation of our favorite algorithm, where we start with a
truly time symmetric choice of time step, with the restrictions that
we only allow changes of a factor two in the direction of increasing
and decreasing the time step, and that we only allow an increase of
time step on the so-called even time boundaries.  This is the approach
given in section \ref{subsec:secondtry}.

In figures \ref{fig1} and \ref{fig2} we show the results of
integrating our highly eccentric binary with these four integration
schemes.  In each case, the largest errors are produced by algorithm
1), smaller errors are produced by algorithm 2), and even smaller
errors appear with algorithm 3).  Finally, algorithm 0) gives the
smallest errors.

Figure \ref{fig1} shows the energy error in the two-body integration
as a function of time.  As is generally the case for time-symmetric
integration, the errors that occur during one orbit are far larger
than the systematic error that is generated during a full orbit.  To
bring this out more clearly, figure \ref{fig2} shows the error only
one time per orbit, at apocenter, the point in the orbit where the two
particles are separated furthest from each other, and the error is the
smallest.

Finally, figure \ref{fig3} shows the same data as figure \ref{fig2},
but for a period of time that is ten times longer.  In both figures
\ref{fig2} and \ref{fig3}, it is clear that the first two block time
steps algorithms, 1) and 2), both show a linear drift in energy.  This
is a clear sign of the fact that they violate time symmetry.  Note
that in both figures algorithm 3) gives rise to a time dependency that
looks like a random walk.  This may well be the best that can be done
with block time steps, when we require time symmetry.

\section{N-body implementation}

So far, we have discussed the implementation of our block-symmetric
algorithm for individual time steps in the case of the two-body
problem.  Of course, for $N=2$, it is not really necessary to use
block time steps, nor is it useful to introduce individual time
steps. The reason we made both of these extensions was to implement
and test our basic ideas in the simplest case.  In this section, we
describe and test our algorithm for the general $N$-body problem.

\subsection{Divide and Conquer: the Concept of an Era}

The major conceptual difficulty in designing a time-symmetric block step
scheme is the global context information that is needed, with extensions
toward the future as well as the past.  In order to determine the
time step for particle $i$ at time $t$, we need the information of all
other particles $j$ at that time.  In general, there will be at least
some $j$ values for which the position and velocity of particle $j$ are
not given at time $t$, because particle $j$ has a time step larger
than particle $i$ {\it and} time $t$ happens to fall within the
duration of one time step for particle $j$.

In such a case, the time step of particle $i$ at time $t$ depends on
the positions and velocities of other particles $j$, that can only be
determined from time symmetric interpolation between the positions
and velocities of each particle $j$ at times earlier and later than $t$.
However, the future $j$ positions and velocities depend in turn on the
orbit of particle $i$, and thus on the time step of particle $i$ at
time $t$.  In other words, there is a circular dependence between the
future positions and velocities of particles $j$ and the time step of
particle $i$.

To make things worse, each of the future positions and velocities of
any of the particles in turn will depend on information that is given
even further in the future.  If we continue this logic, we would have
to know the complete future of a whole simulation, before we could
attempt to time symmetrize that whole history.  And while any
simulation will stop at a finite time, so the number of time steps for
each particle will be a finite number, it is clearly unpractical to
let the very first time step depends on the positions and velocities of
the particles at the very end of simulation.

A more practical solution is to impose a maximum size for any
time step, as $\Delta t_{max}$. If we start the simulation at time
$t_0$, we know that all particles will reach time $t_1 = t_0 + \Delta
t_{max}$, by making one or more steps.  At that time, all particles
will be synchronized.  This means that we can focus on time symmetric
orbit integration for all particles during the interval $[t_0, t_1]$,
without the need for any information about any particle at any time
$t > t_1$.

In other words, we divide and conquer: we split the total history of
our simulation into a number of smaller periods, which we call eras.
Each era extends a period in time equal to the largest allowed time
step $\Delta t_{max}$, or to an integer multiple of $\Delta t_{max}$,
whatever turns out to be the most convenient.

\subsection{Era-Based Iteration}

Let the beginning of a single era be $t_0$ and the end $t_1$.  As we
saw above, in order to obtain the time step size for particle $i$ at
time $t$ within our era, we typically need to know the positions and
velocities of some other particles at times larger than $t$.  The
simplest way to provide this future information is through iteration.

First we just perform standard forward integration, with the usual
kind of non-time-symmetric block step algorithm, for the complete
duration of our era ($t_0 < t< t_1$).  While simultaneously
integrating the orbits of all particles, we store the positions and
velocities (and if necessary higher order time derivatives) for
{\it all} time steps for {\it all} particles during our era.  This will
then allow us to obtain the position and velocity of any particle at
any arbitrary time through interpolation, to the accuracy given by
this first try, which will function as our zeroth iteration.

Next we make our first iteration.  We again perform
orbit integration for our complete $N$-body system, during
$t_0 < t< t_1$.  However, there are two differences with respect to
the first try.  First of all, in order to calculate the force from
particle $j$ on particle $i$, we no longer {\it extrapolate} the orbit
of particle $j$ to the time requested by $i$, but instead we {\it
interpolate} the position and velocity of particle $j$ to the
requested time, using stored positions and velocities of particle $j$
at slightly earlier and later times, using a time symmetric
interpolation scheme.  Secondly, we can now begin to symmetrize the
time step for particle $i$, in the same way as we did it for the
two-body problem in the previous section, with one exception:
we now obtain the estimated time step size at the beginning of the
time step from the current iteration, and we obtain the estimated
time step size at the end of the time step from the previous iteration
(in the two-body case the iteration was done separately for each step).

Subsequent attempts, as second and higher iterations, repeat the same
steps as the first iteration.

As before, we have implemented our iteratively time symmetric block step
algorithm using the leapfrog algorithm as our basic integrator.
Generalizations to higher-order schemes are somewhat more complex, but
follow the same basic logic we are outlining in the current paper.
We have adopted the following time step criterion for particle $i$:
\begin{equation}
\label{eq:nbdt}
\delta t_i = \eta \ {\rm min}_j \frac{|r_{ij}|}{ |v_{ij}|},
\end{equation}
where $\eta$ is a constant parameter and $r_{ij}$ and $v_{ij}$ are
the relative position and velocity between particles $i$ and $j$. To
symmetrize the time step, we simply require that the step sizes that
would be determined by the above criterion at both ends of the time
step are not smaller than the actual time step used.  In other words,
we take the minimum of the two time step values that our criterion
gives us at the beginning and at the end of the time step; this minimum
is our symmetrized time step.  We could have taken the average, but
here we have used the minimum value, for simplicity.

\subsection{Numerical Results for $N=100$ and $N=512$}

\begin{figure}[ht]
\centering
\includegraphics[width=0.90\textwidth]{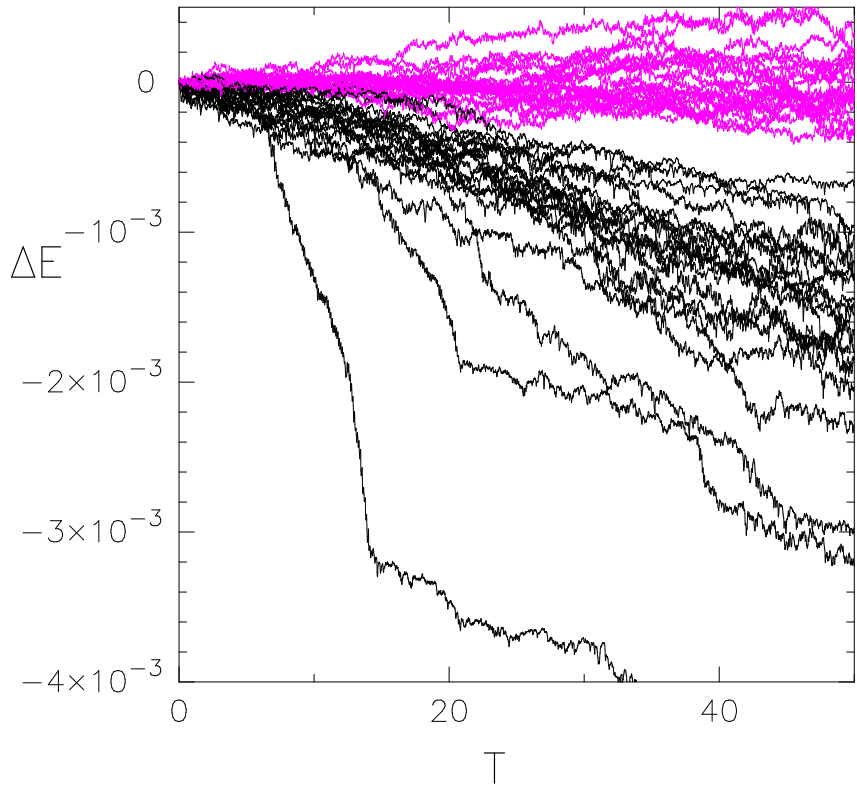}
\caption{Growth of the relative energy error for 100-body runs, starting from
twenty different sets of initial conditions.  For each set of initial
conditions, two integrations have been performed, one without and one
with time-symmetrization (in the latter case, using six iterations).
The twenty lines with time symmetrization form the horizontal bundle
which is slowly spreading in square-root-of-time fashion like a random
walk;  the twenty lines without time symmetrization all show a systematic,
near-linear decrease in energy.}
\label{fig4}
\end{figure}

Figure \ref{fig4} show how the energy errors grow in the 100-body problem.
In each case, we started with random realizations of a Plummer model, where
we used standard $N$-body units, in which the gravitational constant $G=1$,
the total mass $M=1$ and the total energy is $E_{tot}=-1/4$.  We have
integrated each system for 50 time units, with a maximum time step of $1/64$
and $\eta=0.1$ (see Eq. \ref{eq:nbdt}).  We have used standard Plummer type
softening with softening length $\epsilon = 0.01$.  We have carried out forty
time integrations, starting from twenty different realizations of the 
Plummer model.  For each realization, we have integrated the system once
without any time symmetrization and once with time symmetrization using
six iterations to guarantee sufficient convergence.  In our experience,
at least three iterations were necessary to achieve high accuracy.

It is clear that all runs without time symmetry show a systematic drift in
energy, while no such systematic tendency is visible for time symmetrized
runs.  Among the twenty non-symmetrized runs, even the best result
came out worse than the worst result among the symmetrized runs.  We
conclude that time symmetrization can significantly improve the long-term
accuracy of $N$-body simulations.

\begin{figure}[ht]
\centering
\includegraphics[width=0.90\textwidth]{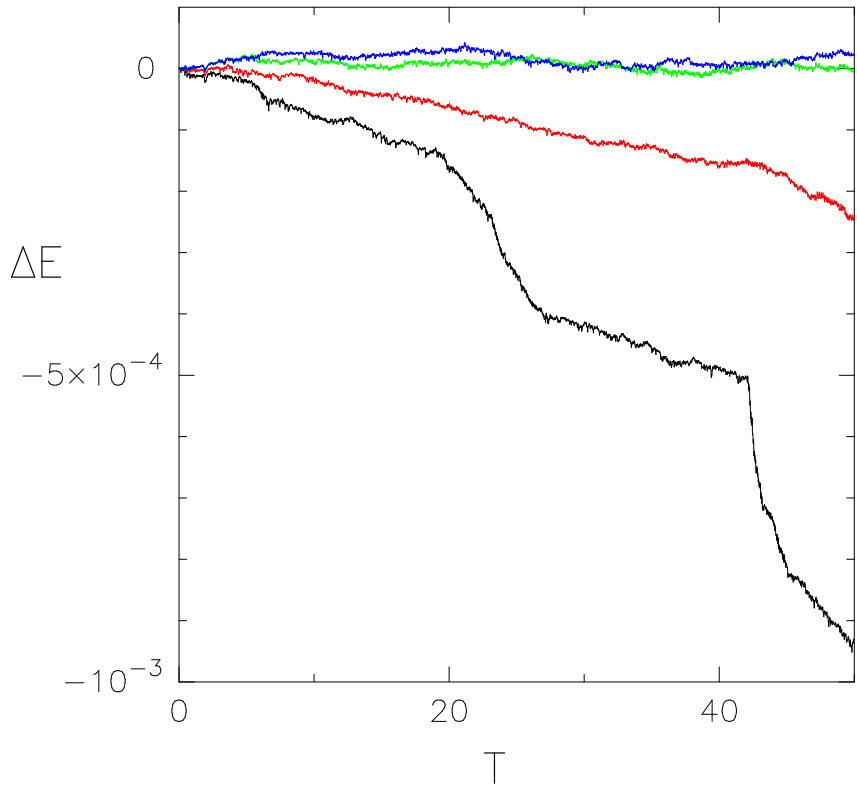}
\caption{Growth of the relative energy error for 512-body runs, starting from
a single set of initial conditions, but using a different number of
iterations.  The lowest curve presents an integration without time
symmetrization.  The curve above that presents the result of time
symmetrization using only one iteration.  The next two curves show
the results of using three and two iterations, respectively; initially,
the third iteration curve rises a bit above the second iteration curve.}
\label{fig5}
\end{figure}

\begin{figure}[ht]
\centering
\includegraphics[width=0.90\textwidth]{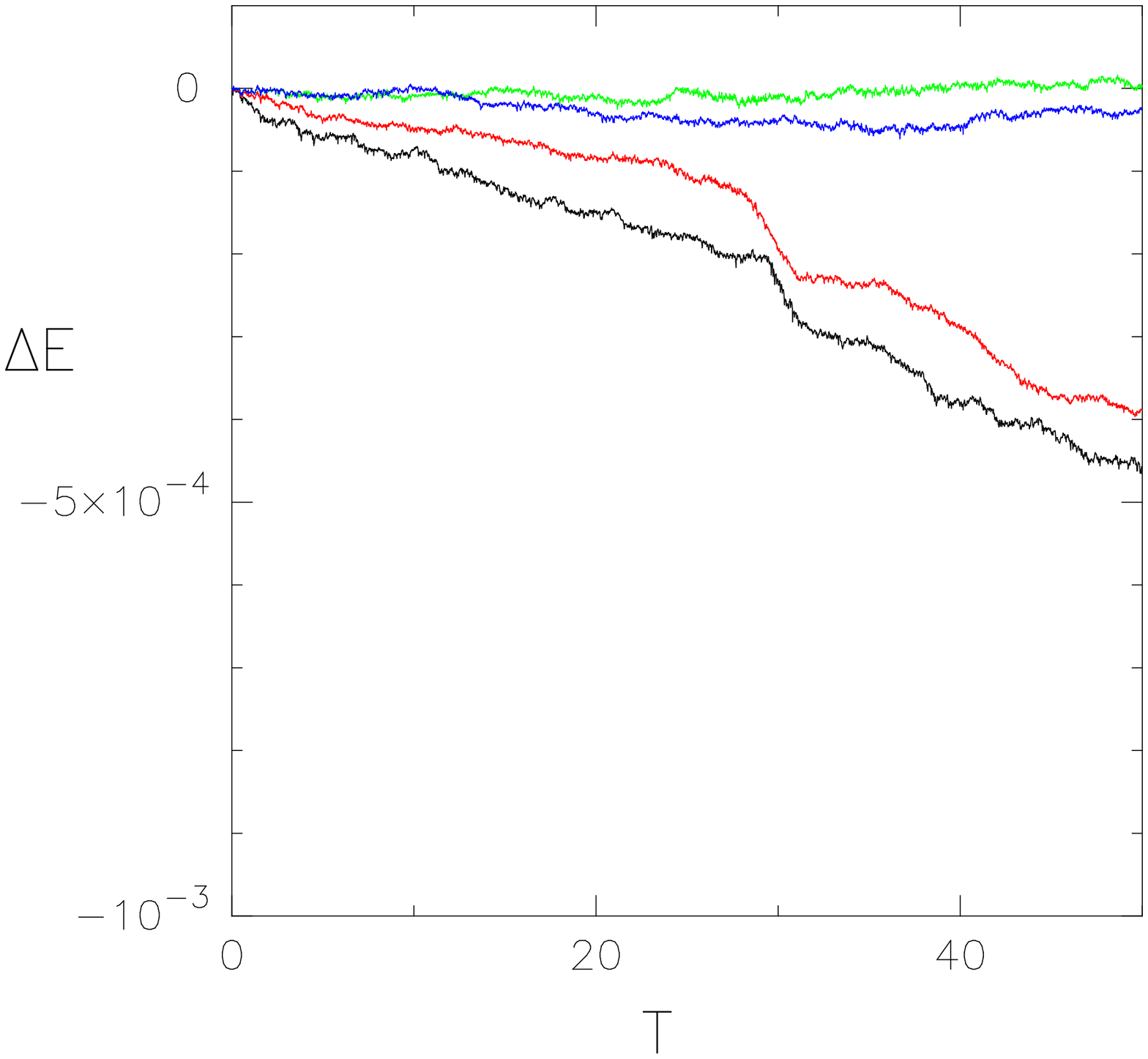}
\caption{Growth of the relative energy error for 512-body runs, like figure 5,
but starting from a different set of initial conditions.  As before, the
lowest curve presents an integration without time symmetrization, and 
the curve above that presents the result of time symmetrization using
only one iteration.  The second iteration curve is the one that stays
above the third iteration curve for most of the run depicted here.}
\label{fig6}
\end{figure}

Figure \ref{fig5} shows similar results for 512-body runs.  Here we have
started from a single Plummer model realization, and the curves show the
effect of varying the number of iterations.  In this case, the second 
and third iteration already show a dramatic improvement in the long-time
behavior of the total energy of the system.  The softening used here is
$\epsilon = 1/512$.  All other parameters are the same as
for the 100-body runs.  For the 512-body case, the improvement is
significantly better than it was in the 100-body runs.  Finally,
figure \ref{fig6} shows that the results depicted in figure \ref{fig5}
are generic: starting from a different set of initial conditions
changes the details but not the overall picture, and again the second and
third iterations show dramatic improvements over the original run and
the first iteration.

We offer the following explanation for the increase in accuracy of the
time-symmetric scheme as a function of particle number.  In these
runs, contributions to the error are largely generated by close
encounters between two particles, since the softening we used is
relatively small.  These error contributions are dominated by weak
encounters, since the softening, while small, is not small enough to
make gravitational focusing significant.  The number of weak encounters
that take place during one time unit, within a particle-particle distance
that is less than the inter-particle distance (of order $N^{1/3}$) is
of order $O(N^{4/3})$.  If our time symmetrization succeeds in replacing
the systematic effects of all these encounters by a random effect, the
result will be a shift from a linear to a square root drift, effectively
replacing the $O(N^{4/3})$ dependence by a $O(N^{2/3})$ dependence.  
We conclude that the relative reduction in the total energy error,
due to time symmetrization, grows with $N$, as $N^{2/3}$.

Whether or not time-symmetric integration is to be preferred, depends
on the number of particles and the duration of the integration.  In
the example case shown in figure \ref{fig4}, the energy error is about
a factor of 10 smaller for time-symmetric integration, but the same
effect could be achieved in a computationally cheaper way by reducing
the step size by a factor of $\sqrt{10}$.  For large $N$, however, as
shown already in figure \ref{fig5}, the effect of time symmetry is
more pronounced. 

For very long integrations, time-symmetric integration is clearly
better, since in that case the error grows in random-walk fashion, as
$\sqrt{t}$, while for any non-time-symmetric scheme scheme the error
grows linearly, proportional to $t$.  In the case of a star-cluster
simulation which covers many relaxation timescales, the particles in
the core can go through a very large number of crossing times.  For
example, a simulation of a globular cluster with $10^6$ stars would
need to cover at least $10^5$ half-mass crossing times.  The crossing
timescale in the core is at least a factor of 100 shorter than that of
the half-mass crossing time.  Therefore, particles in the core need to
be followed for as many as $10^7$ local crossing times.  The difference
between the scaling of $\sqrt{t}$ and $t$ produces a improvement of
a factor of roughly $10^{3.5}$ in the energy error, in the case of
time-symmetric integration.  With a second-order scheme, this translates
into a factor of 100 difference in the necessary step size.  Even with
a fourth-order scheme, it would imply a difference of a factor of ten
in the time step.  Clearly, even applying five iterations will produce
a gain of a factor two with respect to the alternative of having to
decrease the time step by a factor of ten.

\subsection{Description of the Algorithm}

Here we describe the algorithm in some more detail.  To enable the
reader to check the precise implementation of our algorithm, we have
made available the computer code used to generate figures 4, 5, and 6, on
our Art of Computational Science web site
\footnote{http://www.ArtCompSci.org/kali/vol/block\_symmetric/ccode/nbody.c}

Consider the integration of the system from time $t_0$ to $t_1$.  We
can assume for simplicity that the zero point in time has been chosen
in such a way as to be compatible with the era size, so that both
$t_0$ and $t_1$ are integer multiples of the time interval $t_1 -t_0$.

In the first pass through this era, we integrate all particles with
the standard block step scheme, without any intention to make the
scheme time symmetric.  In the calculations reported in this paper, we
have used the same predictor-corrector form of the leapfrog algorithm
as mentioned earlier in Eq. \ref{leapfrog}:

\begin{eqnarray}
r_{new} &=& r_{old} + v_{old} \Delta t
 + \frac{1}{2} a_{old} \Delta t^2,            \nonumber\\
v_{new} &=& v_{old} + \frac{1}{2} (a_{old} + a_{new})\Delta t.
\label{leapfrog2}
\end{eqnarray}

As we discussed before, if one particle $i$ wants to step forward in time,
we need to know the positions of all particles $j$ in order to compute
the force that each particle $j$ exerts on particle $i$.  Among the
particles $j$, many may have a larger time step than particle $i$,
so we may have to predict the position for such a particle,
for the time at which particle $i$ wants to make a step.  In addition,
we need to predict the velocity of particle $j$, because the velocity 
difference between particles $i$ and $j$ are used in determining the
time step size, according to Eq. \ref{eq:nbdt}.

The predicted position $r_{p,\;j}$ for particle $j$ is obtained with a 
second-order Taylor expansion, while for the predicted velocity $v_{p,\;j}$
a first-order expansion suffices:

\begin{eqnarray}
r_{p,\;j} &=& r_j + v_j(t - t_j) + \frac{1}{2}a_j(t-t_j)^2,   \nonumber\\
v_{p,\;j} &=& v_j + a_j(t - t_j),
\end{eqnarray}

The integrated positions and velocities for each particle $i$ at each
time step, $r_{new}$ and $v_{new}$ in Eq. \ref{leapfrog2},  are all stored.

In the next iteration, we proceed in the same way, but the
time step is calculated differently.  At time $t$ for particle $i$, let
the time step calculated according to the criterion (\ref{eq:nbdt}) be
$\delta t$, and the time step not exceeding this $\delta t$ and compatible
with the block step criterion be $\Delta t_p$.  But now we would like to
know the time step size that would be required at the end of this step,
at time $t' = t + \Delta t_p$.

There are two possibilities.  If particle $i$ ended a time step at
time $t'$ in the previous iteration, we are in luck, and we can use
the same procedure we used in the two-body case.  To be specific, we
test if the time step according to criterion (\ref{eq:nbdt}) is smaller
than $\Delta t_p$; if so, we halve the value of $\Delta t_p$.
However, if we are not in luck, and we do not have time $t'$ in the
list of times for which particle $i$ was integrated in the previous
iteration, we simply adopt the time step that we obtained looking in
the forward direction.  In that case, we have to wait till a next
iteration, to apply the time symmetrization procedure.  This
occasional miss explains why the iteration procedure often requires
several iterations before accurate time symmetry can be obtained.

To calculate the force at the new time for particle $i$, we use the
positions of the other particles $j$ calculated by interpolation,
based on the previous iteration, in the following way.  The
interpolation itself is done in a straightforward, linear way.
However, since we have a more accurate position at hand for at least
the starting point of each orbit segment, for particle $j$, we may
as well correct the old orbit segment by shifting it rigidly by an
amount equal to the difference between the starting point of the current
and the previous iteration.

To be specific, the interpolated position at time $t$ for particle $j$
is given by 
\begin{equation}
r_p = (1-f)r_s +fr_e + \Delta r_s,
\end{equation}
where $f = (t-t_s)/(t_e-t_s)$, $t_s$ is the largest time in the list
of times for particle $j$ not exceeding $t$ and $t_e$ is the next time
in that list, immediately following $t_s$.  Here both $r_s$ and $r_e$
are obtained from the stored results from the previous iteration.
The correction term $\Delta r_s$ is defined as
\begin{equation}
\Delta r_s = r_{s,new} - r_s,
\end{equation}
where $r_{s,new}$ is the position of particle $j$ at time $t_s$ in the
current iteration.  As in the case described above, sometimes we are
unlucky, and $r_{s,new}$ is not available.  In that case we just set 
$\Delta r_s$ to be zero, postponing further accuracy improvement until
the next iteration.

To make our predictor-corrector form of the leapfrog integration scheme
consistent, we use the same interpolation scheme for the predictor part,
for the particles to be integrated.  For the corrector part, we used the
trapezoidal scheme, as follows:

 \begin{eqnarray}
 v_{c,new} &=& v_{old} + \frac{1}{2}  \Delta t(a_{old} +a_{new}), \\
 r_{c,new} &=& r_{old} + \frac{1}{2}  \Delta t(v_{old} +v_{c,new}).
 \end{eqnarray}

Here, the subscript {\it old} refers to the value at the previous time,
the subscript {\it c,new} refers to the corrected value at the new time,
$a_{old}$ is calculated with the old values for the positions, and
$a_{new}$ is calculated with the predicted values of the positions.
Note that the first corrected quantity that can be computed is $v_{c,new}$,
based on the old and predicted quantities.  After that, we can also compute
the corrected quantity $r_{c,new}$, based on the old quantities and
$v_{c,new}$.

\section{Discussion}

We have succeeded in constructing an algorithm for time symmetrizing
block time steps that does not show a linear growth of energy errors.
As far as we know, this is the first such algorithm that has been
discovered.  We expect this algorithm to have practical value for a
wide range of large-scale parallel N-body simulations.  While we have
illustrated our approach for simplicity with the leapfrog scheme,
all our considerations carry over to higher-order schemes, as long as
the base scheme can be made time-symmetric when iterated to
convergence.  An example of such a scheme is the widely used Hermite
scheme\citep{Makino-1991}.  We plan to discuss such applications 
in a future paper.

A major novelty in our scheme has been the introduction of a time period,
which we can an era, during which all positions and velocities of all
particles are stored in memory.  These values are retained from one
iteration to the next.  We expect that this procedure will have other
advantages as well, in that it prevents sudden surprises to occur.
For example, if a particle will suddenly require a very high speed,
it may approach another particle with a long time step without any
warning.  As another example, a star may undergo a supernova explosion,
something that other particles will normally only notice when there
current time step has finished.  In both cases, after the first iteration
all particles will have access to full knowledge about these unexpected
events, and during the iteration procedure, they can automatically
adapt to the new situation.

Our new scheme promises to be competitive with traditional
non-symmetrized schemes, especially for very long integration times,
in that the same error bounds may be reached using less computer time.
To prove that this will be the case for realistic applications clearly
requires further detailed investigations, beyond the scope of the
current paper.  The memory use of our scheme may seem formidable, and
indeed, when a large value for the era size is chosen, memory use is
increased significantly over traditional schemes.  For those N-body
calculations that are CPU time limited, this may not be much of a
concern.  However, for large-scale cosmological simulations and other
applications for which memory is important, it is possible to choose
an era size in such a way that the memory requirement of the new
scheme is less than twice the memory requirement of non-symmetric
schemes, at a CPU performance penalty of less than a factor two.  Here
the trick is to take an era size close to the harmonic mean of the
time steps of all particles.  In that way, half the computing cost of
the non-symmetric scheme is associated with particles that have time
steps shorter than this era size.  Those particles with natural time
steps longer than this era choice will see their step size shortened
to the era size, but the total increase in time steps will be less
than a factor two.

The reason that our scheme shows a dramatic improvement in accuracy
is time symmetry, which suppresses linear error growth.  The reason
that our scheme is somewhat complicated is purely empirical: all
else failed, in our attempts to try simpler schemes.  Whether our
procedure is the simplest scheme that actually can produce time
symmetric versions of block time step codes is an open question.
It may well be, but we certainly have no mathematical proof.  This
is an interesting question to be pursued further, for theoretical
as well as practical reasons.

We acknowledge conversations with Joachim Stadel at the IPAM workshop
on N-Body Problems in Astrophysics, in April 2005, where he presented
a time symmetric version of the preprint by \citet{Quinn-1997} which is
similar to our first attempt at a solution, described in section
\ref{subsec:firsttry}.  We thank Sverre Aarseth, Mehmet Atakan G\"urkan,
and an anonymous referee for their comments on the manuscript.
M.K. and H.S. acknowledge research support from ITU-Sun CEAET grant
5009-2003-03.  P.H. thanks Prof. Ninomiya for his kind hospitality at
the Yukawa Institute at Kyoto University, through the Grants-in-Aid
for Scientific Research on Priority Areas, number 763, "Dynamics of
Strings and Fields", from the Ministry of Education, Culture, Sports,
Science and Technology, Japan.

\end{document}